\documentclass[]{spie}  

\usepackage{amsmath,amsfonts,amssymb}
\usepackage{graphicx}
\usepackage[colorlinks=true, allcolors=blue]{hyperref}

\title{Design Concept of a Low-Terahertz Imaging Radar}

\author[a]{Amin Aminaei}
\affil[a]{University of California, Davis. Physics Bldg, 1 Shields Ave, Davis, CA 95616, the USA}

\authorinfo{Send correspondence to E-mail: aaminaei@ucdavis.edu}

\begin{document} 
\maketitle

\begin{abstract}
 Terahertz (THz) imaging proved to be an efficient and powerful nondestructive examination (NDE) method for inspection of materials and detection of defects. Here, we present the development of a Low-Terahertz Imaging Radar(L-TIR) which is specifically designed for the evaluation of Fibre-Reinforced Polymer (FRP) composites. L-TIR will be a compact active radar with a laser pulse transmitter and a receiver which operates at the lower THz band (1 THz-2 THz). L-TIR will probe the structure and sub-layers of the FRP composites which makes it a suitable and fast tool to detect the cracks, breaks and deamination at sub-mm resolution.
\end{abstract}


\section{INTRODUCTION}
\label{sec:intro}  
\begin{figure} [ht]
   \begin{center}
   \includegraphics[width=13cm]{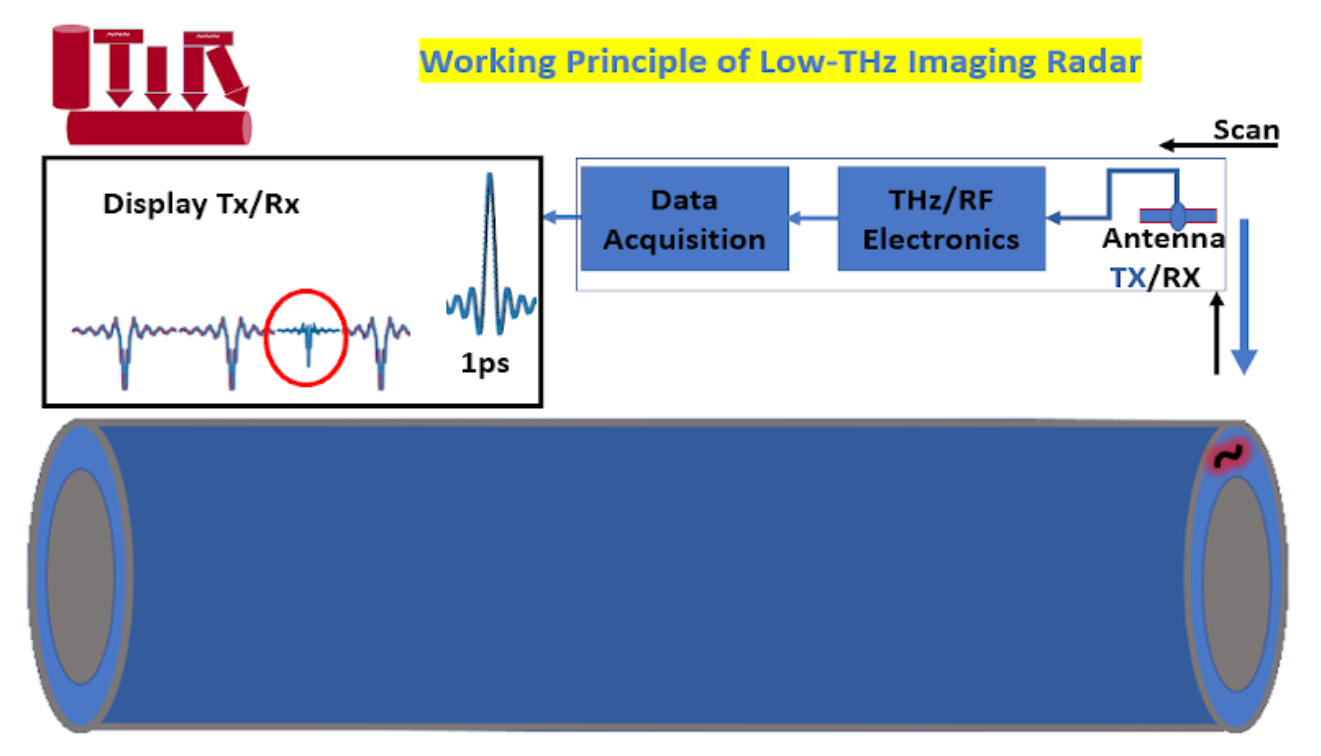}
   \end{center}
   \caption[Fig.1] 
   { \label{fig:fig1} 
An illustration of the working principle of Low-THz Imaging Radar(L-TIR). }
   \end{figure}

THz technology is capable of providng both broad spectrum and high resolution at once. Besides, THz radiation is safe, non-destructive and non-invasive which makes it a suitable tool for many applications. Examples are detection of defects, production inspection, spectroscopy in chemistry and astronomy, material characterization and detection of cancerous tissues \cite{thzapp}. Here, we propose a design concept for a compact portable L-TIR for evaluation and detection of FRP defects. Specifically, it is designed for detection of cracks, delamination and fiber breakage at a resolution of at least 0.3 mm on the water pipes made of FRP composite  \cite{detectme}. Although continuous wave (CW) THz has been also used for detection of defects \cite{cw} , we propose a THz pulsed radar which could generate significant information including the depth of the defects in both time and frequency domain. Working principle of L-TIR is illustrated in Fig.\ref{fig:fig1}.  System includes Tx/Rx antennas, THz/RF electronics and Data Acquisition (DAQ). By transmitting a chain of picosecond (ps) pulses, one can detect possible defects since the reflected echoes from the defects look different. This is due to the discontinuity and non-homogeneous boundaries made by defects. One ps is translated to a wavelength of 0.3 mm which is the minimum resolution requirement for this application. The article continues as follows: in the Sec.\ref{system}, system requirements, the analysis of pulse detection using nominal values is presented then the L-TIR modules and components are introduced. In Sec.\ref{analysis}, analysis and simulation of the TX/RX antennas and filters are presented and the analog electronics and details of digital signal processing (DSP) are discussed. Also, various methods for data interpretation in data post-processing are explained. Finally, a simple layout for a protable L-TIR device is presented in Sec.\ref{ddd}.



\section{SYSTEM REQUIREMENTS}
\label{system}
Pulse characteristics are the key parameters for detection of defects. Based on the defined resolution, 0.3 mm, we examine the feasibility of the detection of deffects by applying the nominal values to the transmitted and reflected pulses. By knowing the characteristics of the pulses, the corresponding L-TIR system including transmitter, receiver, DSP and DAQ modules are designed. 

\subsection{Pulse Detection of Defects}

For the vertical and horizontal resolution, we adapt the formulas which is used for resolution of the Ground Penetrating Radars (GPR)\cite{res}. For the vertical resolution, we use:
\begin{equation}
\label{eq1}
V_{r} = \frac{c.T_{pulse}}{2 \sqrt{\epsilon_{r}}} \, ,
\end{equation}

Where $c$ is the speed of light in vacuum, $T_{pulse}$ is 1 ps and $\epsilon_{r}$ is the relative dielectric permittivity of the FRP composite ($\epsilon_{r} \approx 4$)\cite{eps} . This gives the $V_{r}\approx 0.3 \; mm$ which meets the requirements for detection of the FRP defects.  \\\\ 
For the Horizontal resolution, we use:

\begin{equation}
\label{eq2}
H_{r} = \frac{c}{4.f. \sqrt{\epsilon_{r}}} + \frac{D}{\sqrt{\epsilon_{r}+1}} \, ,
\end{equation}

Where f is the central frequency of the antenna (1.5 THz) and D is the depth of the plane that defects are located. For a nominal value of D = 20 mm , one can obtain $H_{r} 
\approx 9 mm $ which is the minimum distance that two nearby defects at 20 mm depth can be distinguished. \\ 
An example of L-TIR pulse analysis for detection of the FRP defects is shown in Fig.\ref{fig:fig3}. Depending on the power consumption and availability of commercial pulse generators, the Pulse Repetition Frequency (PRF) could be selected in kHz and MHz bands. For a sample of FRP composite, it is assumed that the scanner (i.e. THz antenna) is 5 cm above the sample with a thickness of 25 mm covered with a thin layer of mud/biofouling or debris (5 mm). The timing of each echo pulse is calculated based on the distance of each layer (two\textendash way path) from the antenna and dielectric permittivity of the layer. Distance D, which is the location of defect can be identified by measuring
T, the time difference between received echos the from FRP material and the defect. For a ps pulse and few cm distance, the timing of echos is in the order of ns which can be resolved by DAQ. Considering the timing of echos and power consumption, an PRF in the order of few MHz is proposed for this application.
\subsection{Link Budget}
\begin{figure} [ht]
   \begin{center}

 \includegraphics[width=15cm]{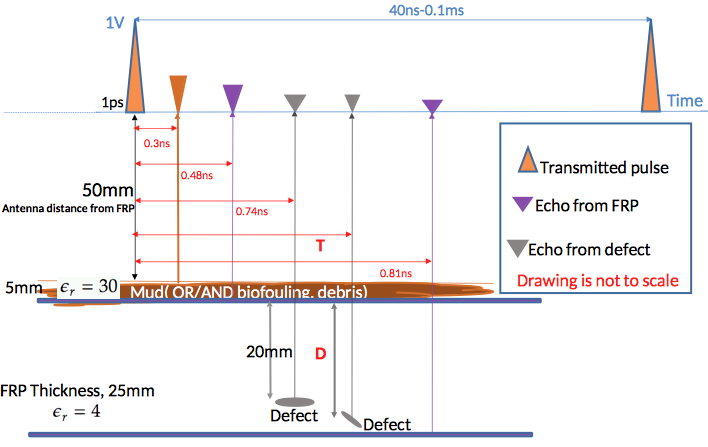}
   
	\end{center}
   \caption[Fig.2] 
   { \label{fig:fig3} 
An example of L-TIR pulse analysis for detection of FRP defects. Distance D can be identified by measuring T, the time difference between received echos from FRP material and defect. For the details see Sec.\ref{system}.}
   \end{figure}

 With a nominal 1 V pulse corresponds to 20 mW (with 50 $\Omega$ impedance), we take a conservative value of 1 mW for radiated power of transmitter antenna. Here we apply following values in Radar equation \cite{radar} to calculate the received power at the receiver: $f_{c}=1 \; THz$, $RF B.W. = 1 \; GHz$, 
 Electrical Conductivity of mud=0.005 S/m, $\epsilon_{r}=30$, 
Max. air attenuation at 1 THz=$10^{5} \; db/km$ \cite{att}, 
Loss tan FRP=0.001,
Antenna Gain TX/RX=7 dBi,
Target= crack with a diameter of 2.5mm,
Total gain (LNA and Power Amp.)=60dB,
Loss (harness connections etc)=10dB,
Transmission and reflection at each border
have been taken into account. \\  
For the example of Fig.\ref{fig:fig3}, the signal power at the receiver before Analog to Digital Converter (ADC) is 0.15 nW and
SNR=15.5 dBW which is high enough to detect the pulse echoes from the background noise(clutter). The certain parameters are summarized in Tab.\ref{tab}. 
 
\begin{table}[ht]
\caption{Certain parameters for calculation of L-TIR Link Budget} 
\label{tab}
\begin{center}       
\begin{tabular}{|l|l|l|l|}
\hline
\rule[-1ex]{0pt}{3.5ex}  Parameter & Value & Parameter & Value  \\
\hline
\rule[-1ex]{0pt}{3.5ex}  $f_{c}$ & 1 THz& RF BW& 1 GHz  \\
\hline
\rule[-1ex]{0pt}{3.5ex}  Antenna Gain Tx, Rx & 7 dBi & Gain of Amplifier(s) & 60 dB \\
\hline
\rule[-1ex]{0pt}{3.5ex}  Total Loss  & 10 dB & SNR & 15 dB \\
\hline
\end{tabular}
\end{center}
\end{table}
\label{sec: pd}

\subsection{L-TIR System Design}
\label{lsd}

L-TIR block diagram including transmitter, receiver, DSP, data storage and display is shown in Fig.\ref{fig:fig2}. Using a THz oscillator and a pulse generator, a chain of ps pulses is generated and transimtted via the transmitter antenna (Tx). The echo pulses are received via a receiver antenna (Rx) which could be the same Tx antenna with a circulator or can be a separate antenna which has been isolated from the Tx antenna by an isolator. The details of the antenna block are explained in Sec.\ref{ant}. The receiver block includes a THz Band-Pass Filter(BPF), a THz amplifier and a THz mixer. Using the THz oscillator, the signal is converted to an RF signal and is filterred by a GHz BPF. An Adjustable Gain Amplifier (AGA) amplifies the RF signal which is then filterred at the desired BW using an RF BPF. The DSP block includes ADC, Field-Programmable Gate Array (FPGA) and Data Distribution Unit (DDU). The FPGA module generates the frequencny spectrum of the digitized data which is stored in the data storage. Using the readout software, data is post-processed and the selected data is shown on the display that can be used to inspect the FRP composite and to detect the possible deffects. The FPGA module also generates pulse trigger for the transmitter(Tx) block. A seperate unit, provides the DC power for all active components. Linear DC converters have a low noise performance and are prefered over the switch mode power supplies. Further analysis and examples of the available components are presented in the next Section (Sec.\ref{analysis}).

\section{SIMULATION AND ANALYSIS }
\label{analysis}   
In this section a basic analysis of the L-TIR modules is presented and examples of the available commercial components are provided.  
\subsection{Antenna}
\label{ant}

Based on the simulation using 4NEC2 software \cite{4NEC2} , we introduce a sub-mm horn antenna with dipole feed where its paraemters are optimized for the L-TIR application. Results of the simulation of L-TIR Tx/Rx antenna are shown in Fig.\ref{fig:fig4}. The horn antenna with diameter of 0.6 mm has a gain of 9.3 dBic at 1.288 THz which has a great impact on the overall system gain to meet the SNR requirements. On the other hand, the corresponding half power (3 dB) beamwidth is roughly $40 ^{\circ}$ which is a fair compromise for fast antenna scanning. 
Of other optimized parameters are the Standing Wave Ratio (SWR) of 1.13 and the antenna efficiency of \%97.2. In Fig.\ref{fig:fig4} the gain variation and electric field at 4 cm distance from the antenna are also shown. Alternatively, commercial THz antennas are available and could be used for the L-TIR. Various types of the THz antennas are introduced in\cite{THzant}. It should be noted that if the same antenna is used for Tx/Rx modes, a circulator (e.g. \cite{circ}) would be needed. For two separate antenna elements, an antenna isolator (e.g. \cite{iso}) would be needed. 

\subsection{Analog Electronics}
The main analog electronics are filters, amplifiers, THz mixer, THz Oscillator and Pulse Generator which are briefly introduced.
\subsubsection{ Filters}
Both THz and RF BPF are required for L-TIR design. For the THz BPF, commercial filters are available at a reasonable price(e.g. \cite{thzfilter}). Commercial RF BPF could be bulky, besides it can be easily designed for a PCB circuit. Fig.\ref{filter} shows a design \cite{rftool} of the 7th Order Elliptic BPF at 1 GHz-2 GHz for the L-TIR Receiver. The desing is matched for a 50 $\Omega$ circuit and is made of the lumped LCs with values in the range of nH , pF and fF.
\subsubsection{ Amplifiers and THz Mixer}
Commercial amplifiers for both THz (e.g. THz Amplifier modules capable of ps pulse amplification in \cite{ap}) and RF modules (e.g. LNA Amplifiers for 1 GHz- 2 GHz in \cite{minic}) are available at small sizes and can be used for the L-TIR design. For THz mixer, the commercial options are limited although the lab prototypes of Superconductor Insulator Superconductor (SIS) mixers have been well developed (e.g. \cite{mixer1} and \cite{mixer2}). For the room temperature mixers, a 3-5 THz harmonic mixer has been developed \cite{rtmixer} which if modified, suits the L-TIR design.  

\subsubsection{ THz Oscillator and Pulse Generator}
Perhaps the greatest challenge of this project is a compact accurate 1 THz oscillator and 1 ps pulse generator which fit a portable device. It seems that the commercial THz oscillators at the right size and frequency might not be currently available. However, resonant-tunneling-diode terahertz oscillators at 1 THz have been fabricated (\cite{osc1} and \cite{osc2}) which could be suitable for the L-TIR. For the pulse generator of the L-TIR, as described in Sec.\ref{system}, 1 ps pulses with PRF of few MHz are needed. The current commercial pulse generators might not have the exact specs. In the literature, a common method for generating ps pulses is step recovery diode (SRD) where the width of pulse can be controlled with a resistor \cite{pg1}, \cite{pg2}. Using the CMOS technology, a pulse width of 2.6 ps with 0.46 mW has been demonstrated \cite{pg3} which seems to be very close to the pulse specs required for the L-TIR. 
\subsection{DSP, Data Storage and Display}
\label{ddd}
\begin{figure} [ht]
   \begin{center}
    

    
 
   \includegraphics[width=15cm]{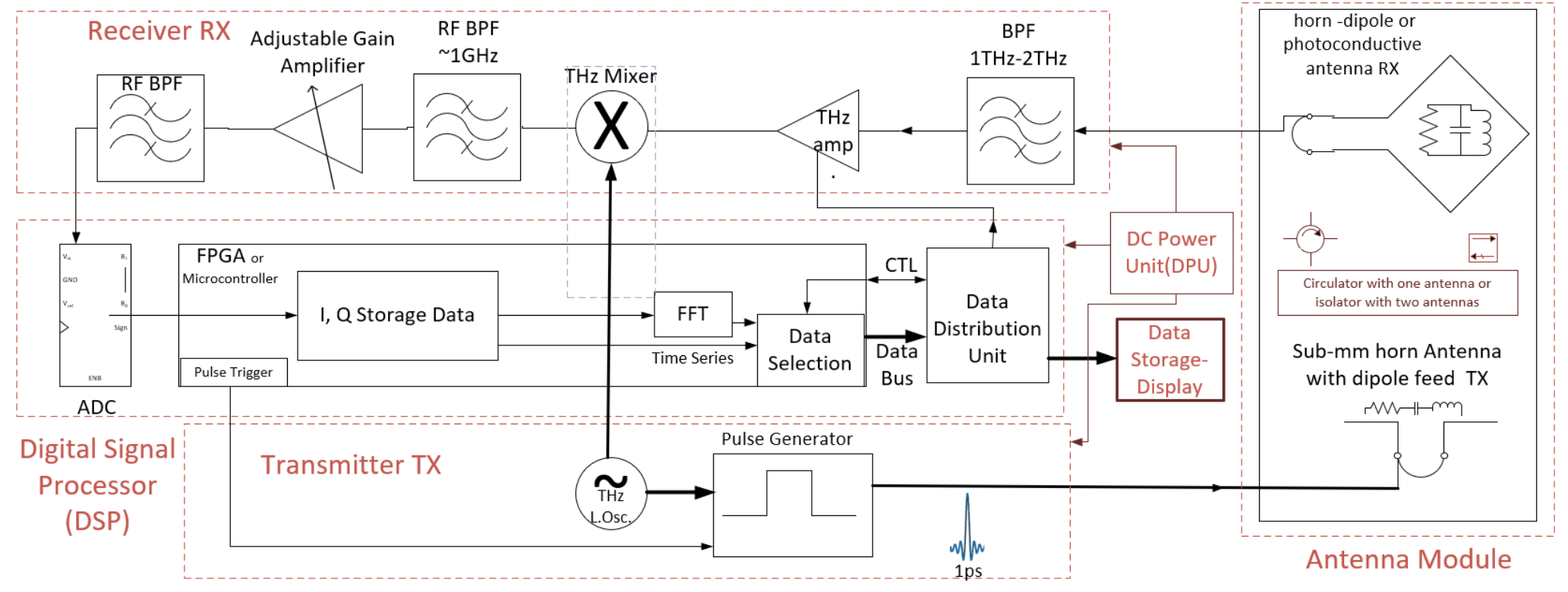}
   
	\end{center}
   \caption[Fig.3] 
   { \label{fig:fig2} 
The L-TIR block diagram described in Sec.\ref{lsd}: Antenna, Transmitter, Receiver, DSP, DC power unit, Data storage and Display. }
   \end{figure}

\begin{figure} [ht]
   \begin{center}
    
 
   \includegraphics[width=15cm]{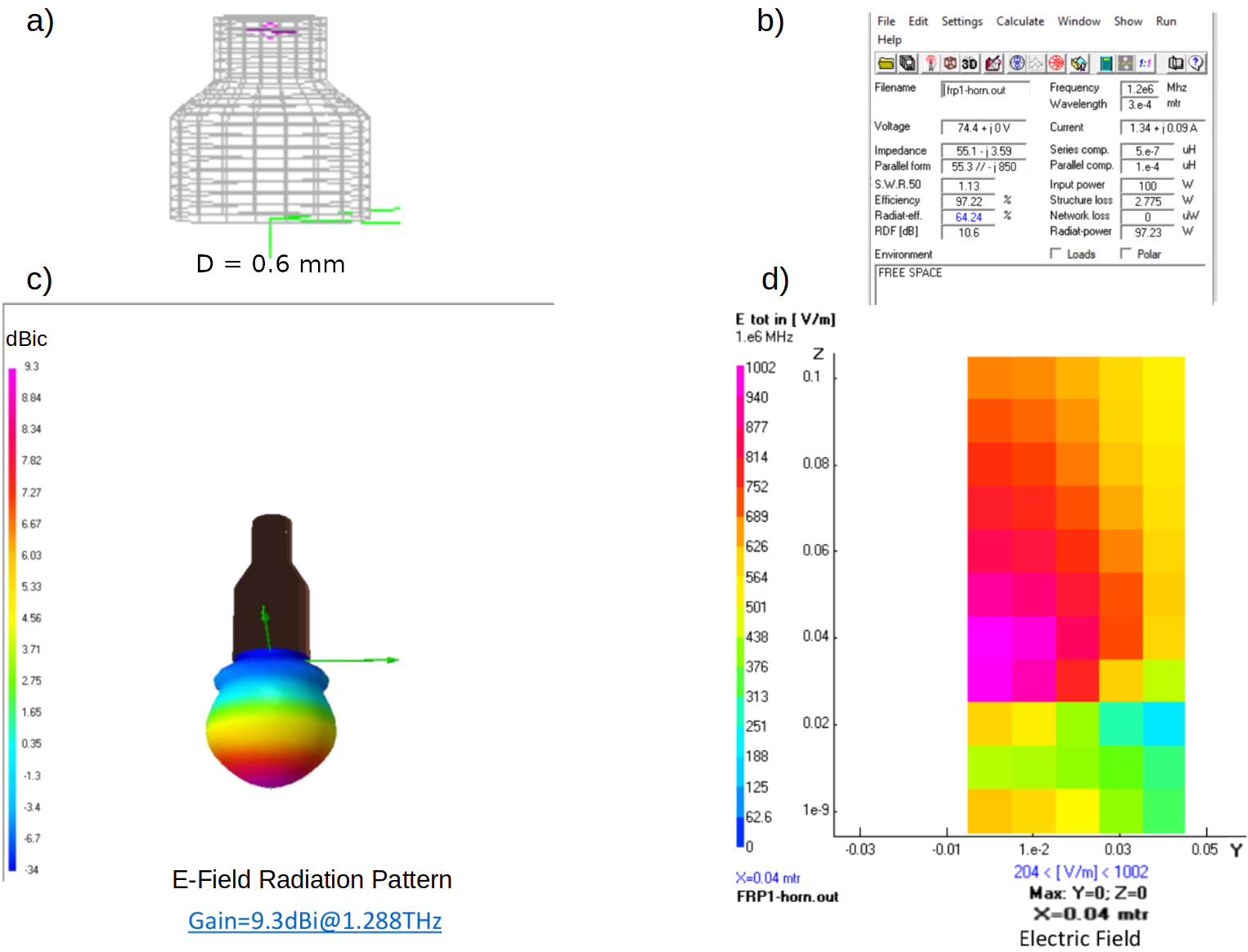}
   
	\end{center}
   \caption[Fig.4] 
   { \label{fig:fig4} 
Simulation of L-TIR Tx/Rx antenna a) Structure of a 0.6 mm horn antenna with dipole feed b) Optimized antenna parameters including VSWR 1.13 and antenna efficiency \%97.2 c) Far E-Field radiation pattern with 9.3 dBic Gain at 1.288 THz d) E-Field at 4 cm distance from the antenna. Simulation is done using 4NEC2 software \cite{4NEC2}.}
   \end{figure}

\begin{figure} [ht]
   \begin{center}
    

 \includegraphics[width=12.5cm]{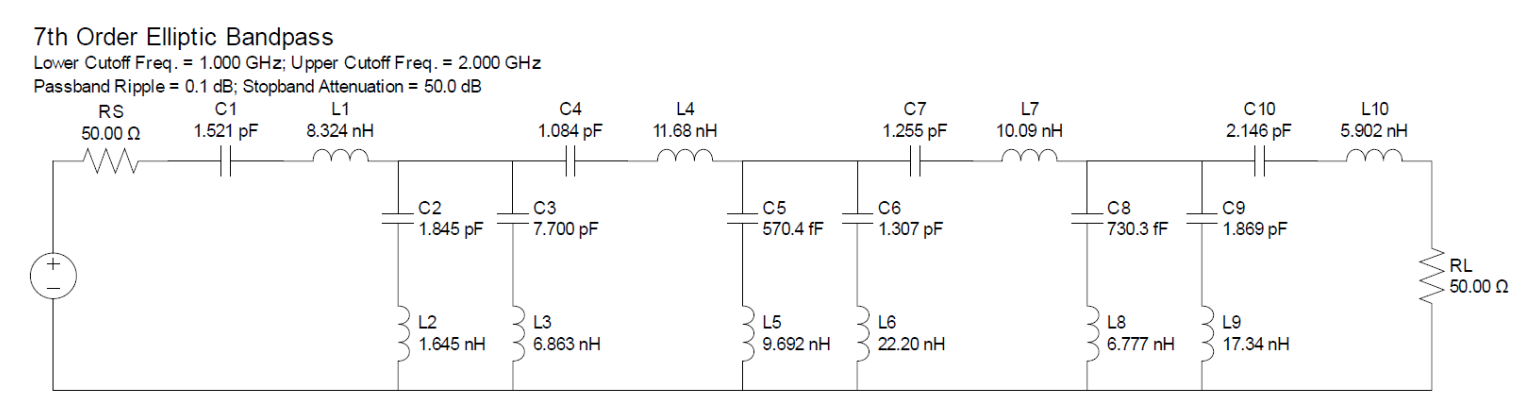}
   
	\end{center}
   \caption[Fig.5] 
   { \label{filter} 

A simulation \cite{rftool} of the 7th Order Elliptic BPF for at 1 GHz-2 GHz for the L-TIR Receiver.}
   \end{figure}

Once the data is transferred to the RF band (1 GHz), it will be digitized and processed in the DSP module. (Fig.\ref{fig:fig2}). For the ADC, sampling rate of more than 2 GHz satisfies the Nyquist criteria. Also, 12-16 bits  would provide a broad dynamic range for the L-TIR. Commercial ADCs such as a TI 12-bit and 6.4 Gsps \cite{adc}  would be suitable. For digital processing, FPGA s are preferred over micro-processors due to the fast and parallel processing. The commercial FPGA kits such as bladeRF 2.0 micro \cite{fpga} would meet the DSP requirements for L-TIR. It covers the frequency range from 47 MHz to 6 GHz, has an AGS and automatic IQ and DC offset correction. It also has a 61.44 MHz MIMO which can be used for pulse triggering with a frequency divider. FPGA generates the frequency spectrum and controls the DDU. As anticipated for the L-TIR layout (Fig.\ref{ant}, the imaging device is planned to be portable (less than 2.3 Kg \cite{portable}) and easy to operate so it excludes the data storage and display. In this regards, a second operator would be needed to monitor the real-time data on a computer display while the first operator scans the  FRP composites.    
\subsection{Data Post-Processing}
For the L-TIR data interpretation, several THz radargrams have been investigated (e.g. \cite{radar1} , \cite{radar2} , \cite{radar3} , \cite{radar4} ). THz imaging can be generated by both Time Domain Spectrometers (TDS) and THz frequency spectrum. For the current L-TIR design, data is taken from the frequency spectrum although in principle TDS can be available as well. It should be noted that the signal is downconverted to the GHz band and need to be rescaled to THz for data interpretation.
The common techniques for clutter removal is required to detect the echo signlas from the defects. Further investigation on L-TIR radargram is subject to the actual data taking from a prototype to distinguish between the echos from the FRP composites and the defects.   
\\\\\\


   
    \begin{figure} [ht]
   \begin{center}
    

 \includegraphics[width=12.5cm]{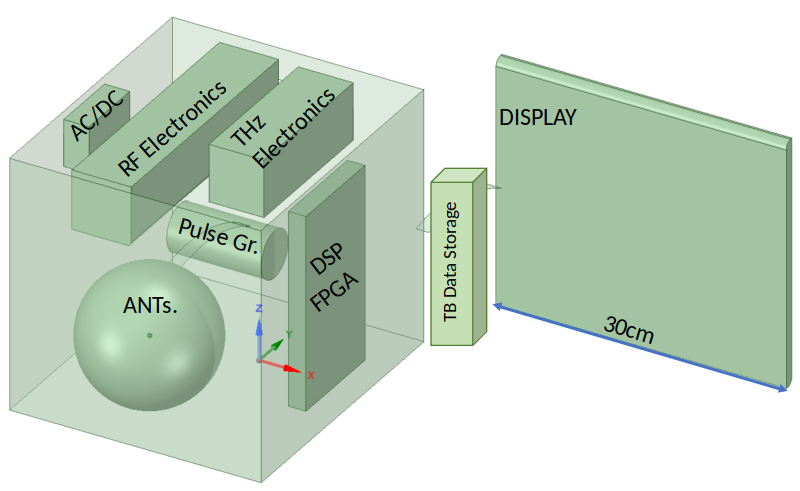}
   
	\end{center}
   \caption[Fig.6] 
   { \label{ant} 
The anticipated layout of the L-TIR. The imaging device excluding the display and data storage is expected to be portable (less than 2.3Kg) and easy to operate.}
   \end{figure}

\clearpage


   \acknowledgments 
The L-TIR design concept was one of the selected proposals for Imperfection Detection challenge organised by HeroX and sponsored by the USA Bureau of Reclamation. We thank sponsors and challenge organiser for their great support. Amin Aminaei is the recipient of the Brinson Prize Fellowship at UC Davis. 

\bibliography{report} 
\bibliographystyle{spiebib} 

\end{document}